# A Brief Summary of EEG Artifact Handling

**Ibrahim Kaya[1]***

[1] Biomedical Engineering Department, University of Miami, Coral Gables, FL 33146, USA

* Correspondence: ixk133@miami.edu;

## Abstract

The applications of Electroencephalogram (EEG) have been extended to out of laboratory and clinics recently due to the advancements in the technical capabilities. There are various advantageous of EEG, making it a preferable method for a wide range of applications; it is a non-invasive method, it is portable, it offers good time resolution and sufficient spatial resolution, besides there are low cost EEG systems available for a commercial use.

Since the early uses of EEG, mainly as monitoring of diseases and pathologies, sleep staging and event related potential researches, it has been intertwined with undesired signal types which we call as artifacts. These pose great challenges in the practice of EEG based methods such as averaging for monitoring and diagnosis of diseases, and single-trial signal analysis for a relatively recent application in brain-computer interfaces. However, many techniques have been developed and under study for better detection and mitigation of these adverse events. The main artifact types and their handling are discussed in this brief summary.

## 1. Introduction

Signal is defined as a function that conveys information about the behavior or attributes of some phenomenon by Priemer (1991). That brings up the question of what the information is. Information can be anything, it can be color, sound, taste, quantity, length, word and so on. An information carrying waveform can have multiple overlapping information in the same space-time.

The signal of interest in the waveform is subjective, it can be color for one and it can be shape for the other. In electrophysiology the waveform under inspection is composed of signal of interest and noise. While signal can be electrocardiography (ECG), EEG or any other physiological signal, noise can be comprised of any unwanted wave source.

## 2. Background

Artifacts are defined as undesired signals that may introduce changes in the measurements and affect the signal of interest by Urigüen and Garcia-Zapirain (2015). EEG can be contaminated in frequency or time domain by artifacts that are resulted from internal sources of physiologic activities and movement of the subject and/or external sources of environmental interferences, equipment, movement of electrodes and cables (Islam et al., 2016). Artifact types and sources are listed in the Table 2-1. External artifacts can be prevented by proper shielding and grounding cables, isolating and moving cables away from recording sites since they act as antennas during operation. On the other hand internal or physiological artifacts are challenging for researchers. The most important artifacts in a typical EEG recording are ocular artifacts or EOG, and muscular artifacts (EMG).

| Artifact | Type | Source |
| --- | --- | --- |
| Eye blink | Ocular | Internal/Physiological |
| Eye movement | Ocular | Internal/Physiological |
| REM Sleep | Ocular | Internal/Physiological |
| Scalp contractions | Muscle | Internal/Physiological |
| Glossokinetic artifact | Muscle | Internal/Physiological |
| Chewing | Muscle | Internal/Physiological |
| Talking | Muscle | Internal/Physiological |
| EKG | Cardiac | Internal/Physiological |
| Swallowing | Muscle | Internal/Physiological |
| Respiration | Respiratory | Internal/Physiological |
| Galvanic Skin Response | Skin | Internal/Physiological |
| Sweating | Skin | Internal/Physiological |
| Electrode movement | Instrumental | External/Extra-physiological |
| Electrode Impedence Imbalance | Instrumental | External/Extra-physiological |

| Cable movement | Instrumental | External/Extra-physiological |
| Electromagnetic coupling | Electromagnetic | External/Extra-physiological |
| Powerline | Electrical | External/Extra-physiological |
| Head movement | Movement | External/Extra-physiological |
| Body movement | Movement | External/Extra-physiological |
| Limbs movement | Movement | External/Extra-physiological |

**Table 2-1:** EEG artifact types and sources from (Islam et al., 2016), Sazgar and Young 2019.

## 2.1. EOG Artifacts

Electrical potentials which are due to eye movements and blinks propagate over the scalp and create hostile electrooculography (EOG) artifacts in the recorded electroencephalogram (EEG). Eye movements are a major source of contamination of EEG. The origin of this contamination is disputable. Cornea-retinal dipole movement, retinal dipole movement and eyelid movement are the three main proposed causes of the eye movement related voltage potential (Croft and Barry, 2000). The direction of eye movements affect the shape of the EOG waveform while a square-like EOG wave is produced by vertical eye movements, blinks leads to a spike-shaped waveform (Vigon et al., 2000) . Blinks which are attributable to the eyelid moving over the cornea, occurring at intervals of 1-10s generate a characteristic brief potential of between 0.2s and 0.4s duration due to eyelid movement over cornea (Barry and Jones, 1965; Matsuo et al., 1975). The blinking artifact generally has an amplitude much larger than that of the background EEG (Croft and Barry, 2000). It is advantageous to have a reference EOG channel during EEG recording in the cancellation of ocular artifact from EEG activity (Urigüen and Garcia-Zapirain, 2015) .

## 2.2. EMG Artifacts

Electrical activity on the body surface due to the contracting muscles are recorded via Electromyogram (EMG) (Urigüen and Garcia-Zapirain 2015). Since independent myogenic activities of head, face and neck muscles are conducted through the entire scalp, it can be monitored in the EMG (Goncharova et al., 2003; McMenamin et al., 2010) The amplitude of this type of artifact is dependent on the type of muscle and the degree of tension (Sweeney et al., 2012; Urigüen and Garcia-Zapirain, 2015). The frequency range of EMG activity is wide, being maximal at frequencies higher than 30 Hz (Anderer et al., 1999 ; McFarland et al., 1997).

### 2.3. Artifact Handling

Artifact avoidance, artifact rejection, manual rejection, automatic rejection and artifact removal are the methods to deal with artifacts (Fatourechi et al., 2007 ). Although it seems a simple solution to cancel EOG and EMG artifacts by instructing subject to avoid blinking or movement, it can result in change of amplitudes in evoked potentials as well as the additional cognitive load (Fatourechi et al., 2007; Ochoa and Polich, 2000; Verleger, 1991). Artifact rejection or manual rejection may require a person dedicated to this purpose of eliminating artifacts one by one in an EEG. The artifact detection by an expert may be subjective, tedious and time consuming, in addition it can't be applicable to online removal (Urigüen and Garcia-Zapirain, 2015). However automatic rejection can automate this artifact rejection procedure while it can eliminate non artifact signals if not properly tuned. The automatic rejection of artifact containing EEG can depend on artifact amplitude based or EEG segment rms based artifact detection and rejection. An example of a simple EEG blinking artifact removal is depicted in Figure 2-1. Since ocular blinks have low frequency content compared to EEG, by low pass filtering, EEG can be reduced while blink artifact still remains at a high voltage level. Thus an amplitude threshold based artifact rejection can be applied. As seen from Figure 2-1, red traces are the EEG and blue are the low pass filtered EEG signal. While a simple artifact rejection (without low pass filtering) using a threshold of 20μV will produce false positives( red traces over 20μV) , in the low pass filtered EEG these false positives are prevented.

Usually one or two channel is dedicated to detect EOG artifacts. There are two widely used procedures for EOG artifacts, EOG rejection where EEG trials with EOG artifacts having VEOG greater than a preset threshold are omitted and EOG correction where the effect of eye movement is tried to be removed from EEG (Croft and Barry, 2000 ).

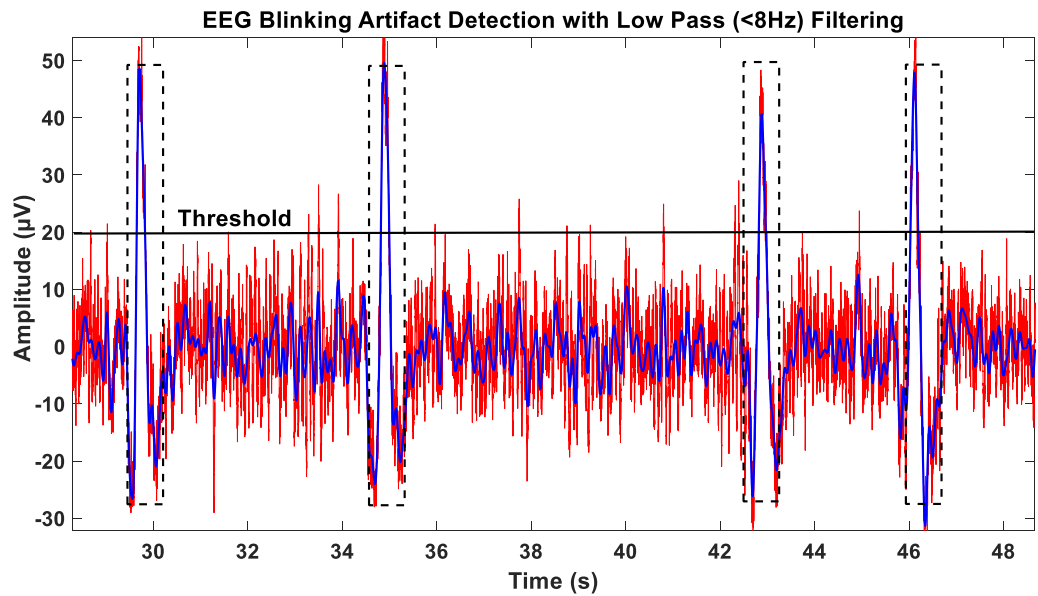

*Figure 2-1* Low pass filtering based EEG blink rejection. Red is raw EEG, blue is low pass filered EEG with 6th order Butteworth low pass filter at 8Hz cut off. The detected artifact containing EEG epochs are shown in dashed rectangles.

Artifacts can distort EEG and the electrophysiologists or physicians can be misled in their clinical interpretation (Hagemann et al., 2001). This makes artifact removal critical in the preprocessing phase prior to analysis. There are many methods to remove artifacts such as Artifactual Segment Rejection, Filtering, Wiener filtering , Adaptive Filtering, Time-Frequency Representation, Wavelet Transform , Discrete Wavelet Transform (DWT), Adaptive Noise Cancellation (ANC) , Wavelet Packet Transform (WPT), Kalman Filtering, Linear Regression, Blind Source Separation (Principal Component Analysis (PCA), Independent Component Analysis (ICA), Canonical Correlation Analysis (CCA) , Minor Components Analysis (MCA)), Source Decomposition, Empirical Mode Decomposition (EMD), Support Vector Machine (SVM), and hybrid methods (Chavez et al., 2018; Chen et al., 2016; Fatourechi et al., 2007; Islam et al., 2016; Lins et al., 1993; Minguillon et al., 2017; Shao et al., 2009; Urigüen and Garcia-Zapirain, 2015, Jiang et al., 2019). A functional dedicated artifact channel which provides complementary aid to identify ECG/EOG is required to remove ocular or cardiac artifacts in the most available methods. (Islam et al., 2016) .

Regression is a common and well established technique in artifact removal, yet it cannot be used to remove muscle noise or line noise, since these type of artifacts have no reference channels (Jung et al., 1998). Having a good regressor (e.g., an EOG) is critical in both time and frequency domain regression methods. It is an inherent weakness that eye movements and EEG signals are bidirectional. When unacceptable amount of data are lost in artifact rejection, delicate artifact removal methods which will preserve the essential EEG signals while removing artifacts are necessary (Jung et al., 1998). One of the most important artifacts is EOG. EEG regions infected with EOG can be rejected from overall EEG signal with simplest artifact rejection where these portions are detected by EOG channels, however these regions still carry brain signals in addition to ocular artifacts and total rejection or subtraction of EOG from them results in loss of brain data (Vigario, 1997; Barlow, 1979 ; Verleger, 1993 ). On the other hand ICA provides extraction of the eye related signals present in the EOG, and removal of this information or artifact, rather than the complete EOG which still has some brain activity (Vigario, 1997). However, detection and removal of transient artifacts such as head and neck muscle contractions and movement are difficult with ICA (Chang et al., 2018). Moreover, adapting ICA as an online method requires high computational power (Chang et al., 2018). On the other hand, Artifact Subspace Reconstruction (ASR) , which is a powerful automated artifact removal method available for both online real-time and offline, can be applied to prevent transient and large artifacts (Chang et al., 2018, Kothe and

Jung, 2014). It also doesn't require additional channel and cleans the data from artifacts. Urigüen and Garcia-Zapirain, (2015) reviewed artifact removal methods for ocular artifacts and concluded as follows either Revised Aligned-Artifact Average (RAAA) and Second Order Blind Identification (SOBI) methods are suggested.

Another method is filtering in frequency domain. For example, EMG activity of contracting scalp sites can hinder the signals of interest in the EEG recordings during an epileptic seizure (Gotman et al., 1981). It was possible to remove this high frequency content EMG activity from EEG spectra by filtering out signals over 25Hz. Independent Component Analysis (ICA), a Blind Source Separation (BSS) method, is often used to remove EEG artifacts based on statistical approach of spatial filtering and separation of multiple channel EEG data into spatially fixed and temporally independent components (Jung et al., 1998; Jung et al., 2000 ; Grouiller et al., 2007). An advantage of ICA is that it doesn't rely on a reference channel (Jung et al., 1998). However, Many artifact removal algorithms are compared in review by Urigüen and Garcia-Zapirain, (2015) , RAAA, SOBI, and Adaptive Mixture of Independent Component Analyzers (AMICA) are the preferred artifact removal methods for EOG, EMG and ECG artifacts.

Since EEG is widely used as a clinical tool to monitor or diagnose patients, doctors can be misguided in case of artifacts and EEG can be misinterpreted. For this reason, artifact removal becomes a crucial point for some cases such as epilepsy monitoring in an EEG/fMRI recording room. Today EEG and fMRI are two distinct but closely related and complementary methods. While fMRI provides high spatial resolution for localization of phenomena in the brain, EEG on the other hand results in better temporal resolution (Huster et al., 2012; Vanni et al., 2004 ; Wibral et al., 2009, 2010 ). One should be careful about the experiments involving both fMRI and EEG because there are many unwanted electromagnetic sources interfering with EEG. For example, the false identification of spikes are highly possible since residuals of ballistocardiogram (BCG) artifacts have similar shapes as epileptic spikes (de Munck et al., 2013). The factors that can lead to differences in the artifact are linked to the subject and experimental setup, Debener et al. (2009). There are imaging artifacts, cardiac related Ballistocardiogram artifacts (BCG) , EOG and EMG artifacts in an EEG inside MRI (Grouiller et al., 2007). Static field (B0) and the time-varying fields of radio-frequency excitations and of imaging gradients, generate artifacts in the EEG known as ballistocardiogram (BCG) and imaging artifacts (Grouiller et al., 2007; Bonmassar et al., 2002; Allen et al., 2000; Felblinger et al., 1999) The pulse artifact which can be observed in EEGs recorded inside MR scanners easily, is due to a fundamental cause that any movement of electrically conductive muscles in a static magnetic field generates electromagnetic induction and it is proportional to the static field, generally larger at higher field strengths (Debener et al., 2008; Debener et al., 2009). Pulsations of the scalp arteries are the main cause of this type of BCG artifact. (Allen et al., 1998; Ives et al., 1993). The study of Grouiller et al., (2007) compared different imaging artifact removal techniques and various cardiac artifact correction techniques in both simulated EEG data and in real experimental data. They concluded that there is no key for every door, some algorithms work well for some case and others might work well for other cases. Certain algorithms may be preferred depending on the type of data and analysis method (Grouiller et al., 2007). Adaptive Optimal Basis Set (aOBS) is an algorithm to automatically eliminate BCG artifacts yet preserving the neural origin signals in EEG (Marino et al., 2018). It can be used efficiently for simultaneous fMRI and EEG recordings.

Manual artifact detection is still the most common method for artifact handling for sleep stage classification, however, the long time required and the difficulty to apply it to large datasets poses the main disadvantages (Malafeev et al. 2018). Malafeev et al. (2018) tested 12 simple algorithms that are applicable with a single EEG channel for ease of use. It was found that automatic artifact detection in EEG during sleep within large datasets is possible with simple algorithms. Among these, Power thresholding 25–90 Hz (PT25), Power thresholding 45–90 Hz (PT45) and Autoregressive (AR) models had ROC areas above 0.95. In addition, online detection is also possible with the majority of these simple algorithms.

Artifact removal in BCI applications are getting more attention. It was shown by studies that artifacts generated by EOG and EMG activities affect the neurological signals utilized in a BCI system (Goncharova et al., 2003; McFarland et al., 2005).Although there are significant researches into artifact removal for BCIs such as Fully Online and Automated Artifact Removal (FORCe), Lagged Auto-Manual Information Clustering (LAMIC), Fully Automated Statistical Thresholding for EEG artifact Rejection (FASTER) and K-Singular Value Decomposition (K-SVD), the field lacks an effective artifact removal technique for short segments of single channel EEG (Chen et al., 2014; Chen et al., 2016 ; Daly et al., 2013, 2015;   Khatun et al., 2016 ; Nolan et al., 2010; Sreeja et al., 2018; Sweeney et al.; 2012). The surrogate-based artifact removal (SuBAR) technique proposed by Chavez et al. (2018) effectively cancels EOG and EMG artifacts from single-channel EEG. Chang et al. (2016) proposed a method for detection of eye artifact from single prefrontal channel which is useful for headband-type wearable EEG devices with a few frontal EEG channels. Compared to conventional methods the accuracy of detecting ocular artifact contaminated epochs was significantly better. Daily-life EEG-BCIs are getting popular and artifact removal techniques for these BCIs must have some critical features such as; must be performed outdoor, with portable wearable wireless device, with real EEG signals, compatible with daily life tasks, must have simple electrical montage, must use dry electrodes, must remove complex artifacts, must work only EEG without reference, must work online and must work with single electrode channel. More research into artifact removal other than ocular and cardiac artifacts is necessary especially for those daily-life EEG BCIs (Minguillon et al., 2017).

## 3. Conclusion

The number of artifact handling techniques and algorithms are increasing drastically, however the artifact problem is still challenging for many applications. While simple measures such as artifact avoidance and artifact rejection can be utilized in some applications, most of the cases require special methods dedicated to handle artifacts in order to significantly reduce their harmful effects on signal of interest. A generic method for all sorts of artifacts is still missing.